# Quantum Computer Control using Novel, Hybrid Semiconductor-Superconductor Electronics

Document ZF001 v1.05

Erik P. DeBenedictis
Zettaflops, LLC, Albuquerque, NM 87112
erikdebenedictis@gmail.com

*Abstract*—Inspired by recent interest in quantum computing and recent studies of cryo CMOS for control electronics, this paper presents a hybrid semiconductor-superconductor approach for engineering scalable computing systems that operate across the gradient between room temperature and the temperature of a cryogenic payload. Such a hybrid computer architecture would have unique suitability to quantum computers, scalable sensors, and the quantum internet.

The approach is enabled by Cryogenic Adiabatic Transistor Circuits (CATCs), a novel way of using adiabatic circuits to substantially reduce cooling requirements. In a hybrid chip of CATCs and a second technology, such as Josephson junctions (JJs) or cryo CMOS, the CATCs complement the speed, power, and density of the second technology as well as becoming a long-sought cryogenic memory.

This paper describes higher-level design principles for CATC hybrids with a quantum computer control system that includes CATC memory, an FPGA-like logic module that uses CATC for dense configuration logic and JJs for fast configured logic, and I/O subsystems including microwave modulators and low frequency control signals.

## I. INTRODUCTION

This paper shows how CATCs can expand the range of applications of cold electronics and increase the performance of those applications at scale. Fig. 1 shows the structure of a cold electronic system in enough detail to introduce the main points.

In this paper, cold, scalable electronics address computational tasks that include a scalable information processing payload that only functions when cold, such as set of sensors, qubits, or cryogenic computing components.

The payload's behavior is designated $q(N)$, where $q$ is the functional behavior of the payload and $N$ is the number of sensor elements, qubits, the size of the computational problem, or generally the scale factor. The significance of a scalable cryogenic payload is most easily seen in quantum computing, where computer scientists analyze the algebraic expression for $q(N)$ to determine quantum speedup.

Since the payload runs cold, data to and from the payload will be routed across the temperature gradient in steps, illustrated in Fig. 1 with stages at room temperature designated as 300 K, 4 K, and 15 mK.

Computational systems scale up in accordance with Rent's rule, which has been adapted to cryogenic systems of the form illustrated in Fig. 1.[1] However, a key idea in this paper is that the ideas in Ref. 1 are incomplete. Rent's rule is due to an engineer E. F. Rent analyzing integrated circuits, PC boards, and full computer systems, observing the number of external interface wires per internal component decreases as one moves up a system's hierarchy, with the rate of decrease correlating with the system's scalability.

For Fig. 1, this means the intermediate stages will perform electrical or format translation, control loops, and buffer data to reduce bandwidth over external interfaces. Thus, the overall task in Fig. 1 can be designated as $f(N) = (f_{300} \circ f_4 \circ f_{0.015} \circ q)(N)$, where $f_T$ is the functional behavior of the stage running at temperature $T$ and the operator $\circ$ is function composition.

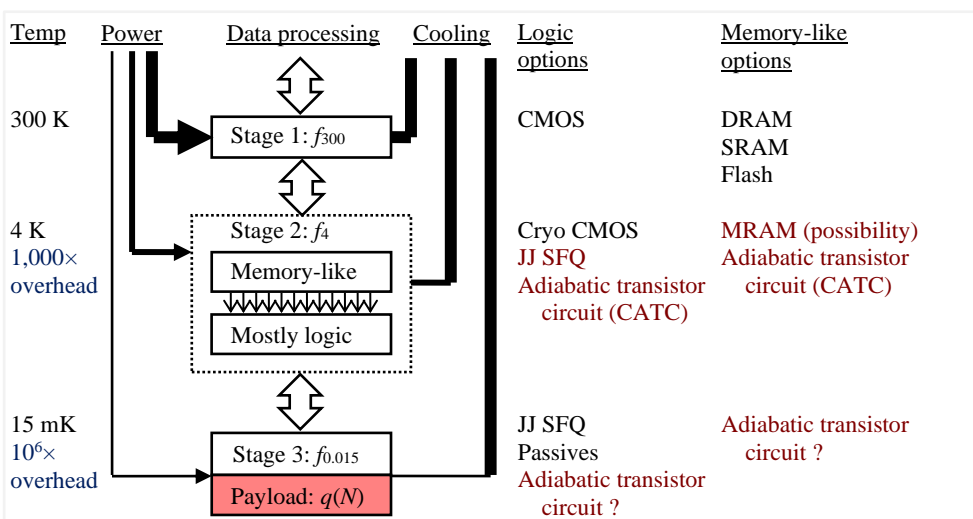

Fig. 1. The overall task is to perform function $f(N) = (f_{300} \circ f_4 \circ f_{0.015} \circ q)(N)$ at any specific scale $N$, possibly including determination of asymptotic scalability as $N \rightarrow \infty$. Data transfer between stages at different temperatures will be accompanied by a proportional amount of parasitic heat and noise flow from the hotter stage to the cooler one.

However, the adaptation of Rent's rule in Ref. 1 does not account for the energy consequences of processing data at a nonstandard temperature $T$. Heat generated at temperature $T$ has to be removed by a refrigerator leading to about 1,000× total energy at $T = 4$ K and 1,000,000× or more at $T = 15$ mK (these overheads are explained in more detail later). Energy efficiency also varies with temperature for many computing technologies.

Thus, achieving scalability in cold electronics also depends on giving the designer a set of devices, circuits, and architectures that are energy efficient at temperatures set by the application's requirements. The right side of Fig. 1 shows some technologies available to designers as a function of $T$.

This paper introduces CATCs as an important technology to address the need above. It has been known for decades that adiabatic transistor circuits can move what is essentially waste energy from logic circuitry to the vicinity of the power supply in hopes that it could be collected and recycled, thereby reducing wall plug power consumption. Unfortunately, energy recycling power supplies were never perfected so today's circuits simply move waste energy from one place to another before it dissipates as heat.

However, if the logic circuity is in the cold environment while the power supply is at room temperature, adiabatic circuitry will move waste energy from a place where heat removal incurs a substantial cooling overhead to a place where it does not. Lessening refrigeration overhead gives about as much benefit as was originally proposed for adiabatic circuits if the collection and recycling of heat into energy could have been perfected.

Moreover, some CATCs are actually logic families that trade speed for energy efficiency over several orders of magnitude, although this flexibility comes with design constraints.

The most straightforward application of CATCs is as the memory-like part of a hybrid with a second technology such as JJs or cryo CMOS, as shown as function $f_4$ in Fig. 1. The term memory-like refers to structures whose purpose is to hold information or state such as shift registers and flip flops in addition to random-access memories that dominate the consumer marketplace. This paper includes a series of architectural structures that essentially use CATCs to buffer data like FPGA configuration strings and digitized waveforms for use by a faster logic technology.

CATCs are clearly applicable to quantum computing. As stated above, computer scientists project quantum speedup from $q(N)$, but the computer will actually perform $(f_{300} \circ f_4 \circ f_{0.015} \circ q)(N)$. The ideas in this paper are intended to give the designer a fair chance to implement $f_{300} \circ f_4 \circ f_{0.015}$ so the quantum speedup will be available to a user at room temperature instead of the control electronics forming a bottleneck that reduces the speedup.

## II. BACKGROUND

People invented electronic computers almost a century ago, apparently assuming that computers would run at the same temperature as the people who invented them. People have been busy scaling up room-temperature computers ever since, but treating applications requiring nonstandard temperatures as special cases because there were apparently not enough of them to justify developing a general set of design principles.

Due to the large size of the computer industry, there are highly refined design processes for just about every conceivable combination of room-temperature computing devices. Laptops and smartphones use CMOS for logic, DRAM for memory, and Flash for storage. Theoretically, a computer could be made of just two or even one of these device types, but the resulting system would be far from optimal because the designer would not have the freedom to implement internal tasks with devices optimized for those tasks.

Cryogenic design processes are not as mature. For logic, the designer has a choice of JJs and cryo CMOS, but there are no good memory options. Furthermore, JJs and cryo CMOS are at extreme ends of a spectrum: Both are about the same speed, but JJs use about 10,000× as much chip area while cryo CMOS uses about 10,000× as much energy per logic operation.

Quantum computers are now in the public eye for potential large-scale applications, with some qubit types requiring operation near absolute zero. To evaluate the scalability of these qubit types will require a structure like the one shown in Fig. 1, but the underlying technology, once developed, would also apply to scalable cryogenic sensor systems, as may be found on a spacecraft in search of extrasolar planets, and proposed cryogenic supercomputers that rely on the higher performance and energy efficiency of JJs.

### A. Device performance and temperature

A central issue is how the energy required for computation varies with temperature. With energy measured in Joules, today's ubiquitous CMOS uses nearly the same amount of energy at any temperature where it works in the first place.[2] However, the resulting heat must be removed to room temperature by a cryogenic refrigeration system. If refrigerators were 100% Carnot efficient, a total of 300 K/$T$ would be drawn from the wall plug and ultimately dissipated into the 300 K environment. Refrigerators are not 100% Carnot efficient, leading to a power multiplier closer to 1,000× for computation at liquid Helium temperatures of 4 K and 1,000,000× or more at typical qubit temperatures of 15 mK.

However, the physical limits of computation, such as Landauer's minimum dissipation,[3] are expressed in entropy units of $kT$, where $k = 1.38 \times 10^{-23}$ is Boltzmann's constant. As the operating temperature $T$ goes down, so does the minimum energy, so one might wonder if actual devices become more energy efficient as they are cooled:

JJ electronics can be engineered for proportionally lower energy down to 15 mK or lower, although the energy per operation does not change once a circuit has been manufactured.

While the dissipation of CMOS is nearly independent of temperature, this is due to the structure of the CMOS circuit. If CMOS transistors are used in an adiabatic circuit, the dissipation can change with temperature, including after the circuit has been manufactured.

The year-over-year improvements in CMOS energy efficiency, sometimes called "Moore's law," slowed in the early 2000s and led to an extensive search for an alternative to the transistor. The scatter plot in Fig. 2a[4, 5, 6] shows the energy and delay for many of the devices considered by international research programs seeking a "Beyond CMOS" device. Fig. 2 implicitly assumes room temperature operation.

There are three additional light-blue solid dots in Fig. 2a for JJ circuit families called Reciprocal Quantum Logic (RQL) and Adiabatic Quantum Flux Parametron (AQFP). The dots with "whiskers" above them show the sum of device dissipation plus the energy of the cryogenic refrigerator that removes the device's heat from 4 K to room temperature. The span of the whisker represents the range of efficiencies found in typical refrigerators. Only points in the blue lasso are applicable to room temperature operation, leading some people say "Moore's law is ending" because none of the devices in the blue lasso are much better than CMOS.

Applications of CATCs include sources and sinks of data at a cryogenic temperature. The devices that directly interface with this data will require cryogenic refrigeration, making the vertical, energy axis in Fig. 2a the pre-refrigeration energy dissipation. So, for the purposes of this paper, all the devices in Fig. 2a incur the same refrigeration overhead, making the JJ circuits in the red lasso stand out because they dissipate much less power to begin with.

### B. Adiabatic circuits

Fig. 2b shows a qualitative difference in the energy per operation of CMOS and an adiabatic transistor circuit. Each of the dots forming the curves are the result of a Spice simulation of a shift register at a different clock rate and with different parameters.

A CMOS circuit's characteristic energy/op sets the level of the flat middle section of its curve. The left end of each CMOS curve is at the clock period where the circuit can no longer charge the wire capacitance within the available time and the circuit stops working. The CMOS curves rise on the right due to leakage, which is significant only at very low frequencies.

Fig. 2a and Fig. 2b are both log-log plots, although the scales are different. The reader will see that Beyond CMOS data in Fig. 2a represents each device by a single point that corresponds to the left end of the horizontal line in Fig. 2b—which is a valid abstraction because the CMOS circuits always create horizontal lines.

However, the adiabatic circuits in Fig. 2b become more energy efficient as the clock slows down; Fig. 2b is based on the 2LAL circuit,[7] but there are many other adiabatic circuits that would create similar curves. The curves should have slope −1 according to circuit equations, and the reader can see that the curves are nearly parallel to the dotted likes of constant energy-delay product.

(a) Beyond CMOS devices

(b) **Energy/op vs. freq., TSMC 0.18, CMOS vs. 2LAL**

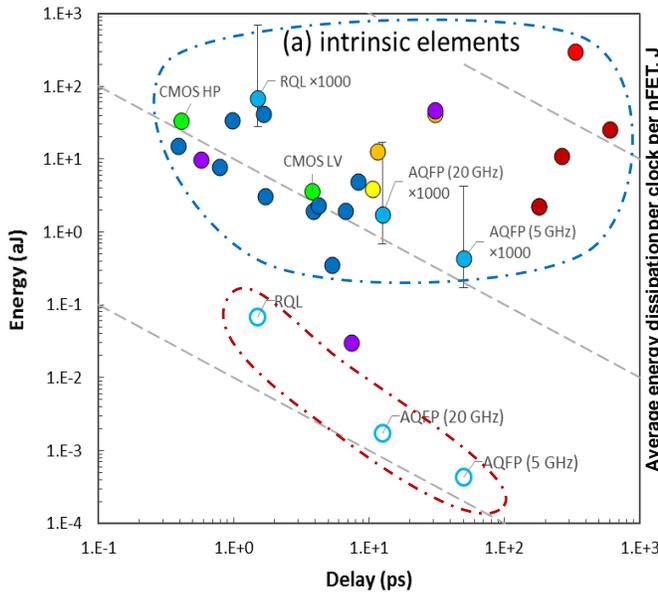
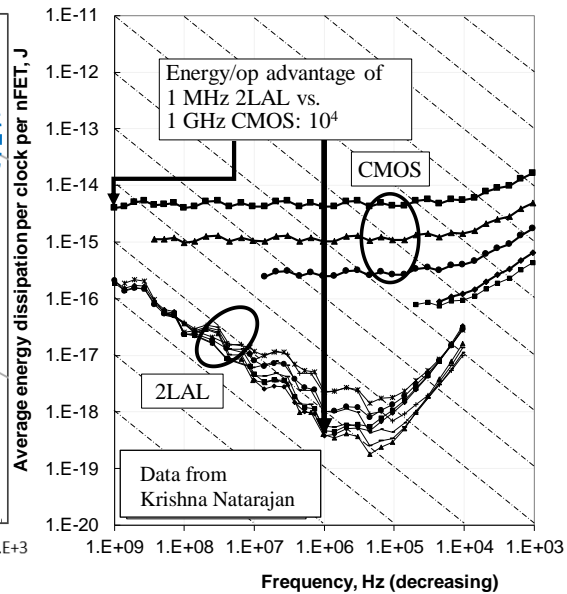

Fig. 2. Left: An energy-delay plot of a comprehensive set of logic devices at room temperature, yet including JJs operating at 4 K in three circuits (RQL and AQFP at two speeds). None of the devices stand out at room temperature because only the superconducting ones have refrigeration overhead. However, at 4 K they all require refrigeration, causing the superconducting devices to stand out. (The outlier purple dot is a BisFET, which is an immature device only demonstrated a mK temperatures.) Right: A plot of many Spice simulations of CMOS and adiabatic (2LAL) circuits, all built from the same transistors, on the same log-log axes. CMOS circuits have unchanging energy/op, leading to a horizontal line, but adiabatic circuits become more efficient at lower frequencies. This report exploits the difference.

The global research community abstracted "Beyond CMOS devices" to the single points in Fig. 2a, obscuring the fact that there are also "Beyond CMOS circuits," such as adiabatic circuits, that have a different behavior.

This paper will show the benefits of these adiabatic circuits.

*C. The 2LAL circuit*

Fig. 3 illustrates operation of 2LAL adiabatic circuits,[7] which is the example that will be used throughout this paper. 2LAL is entirely based on transmission gates, shown in (a). A transmission gate comprises an n-type and a p-type FET connected at their sources and drains. This two-transistor structure acts like an SPST switch, connecting the $A$ and $B$ sides when $P$ is true and an open circuit when $P$ is false. All signals in 2LAL are dual rail, meaning every $A$ is accompanied by a $-A$ elsewhere in the circuit, so the schematic for a transmission gate, shown in (b), comprises two pairs of transistors with inverted signals on the sources and drains.

A 2LAL shift register comprises four repetitions of the structure in (c), the repetitions differing due to the advance of the four-phase clock. There is a logic family built around 2LAL, with (d) illustrating an AND function. Further details on 2LAL logic can be found in Ref. 7.

The schematics (e) and (f) illustrate CMOS and 2LAL systems for easy comparison. CMOS has two DC power supply leads, $V$ and GND, whereas 2LAL has trapezoidal waveforms, $\phi_0$-$\phi_3$, in four leads, which function as both the clock and power supply. Cryogenic implementations typically place the power supply at room temperature, which is the case with the $\phi_0$-$\phi_3$ signal generator. A high-temperature superconductor may be used for the segment of each power supply lead below around 77 K to avoid generating excessive heat in the cold environment.

Since the 2LAL logic family contains a universal gate set, only four wires are required between the room-temperature signal generator and the cryogenic environment irrespective of the complexity of the logic. Of course, additional wires may be needed for I/O and if special voltages are required. Other adiabatic logic families, such as SCRL[8] may use a different number of combined clock and power supply wires.

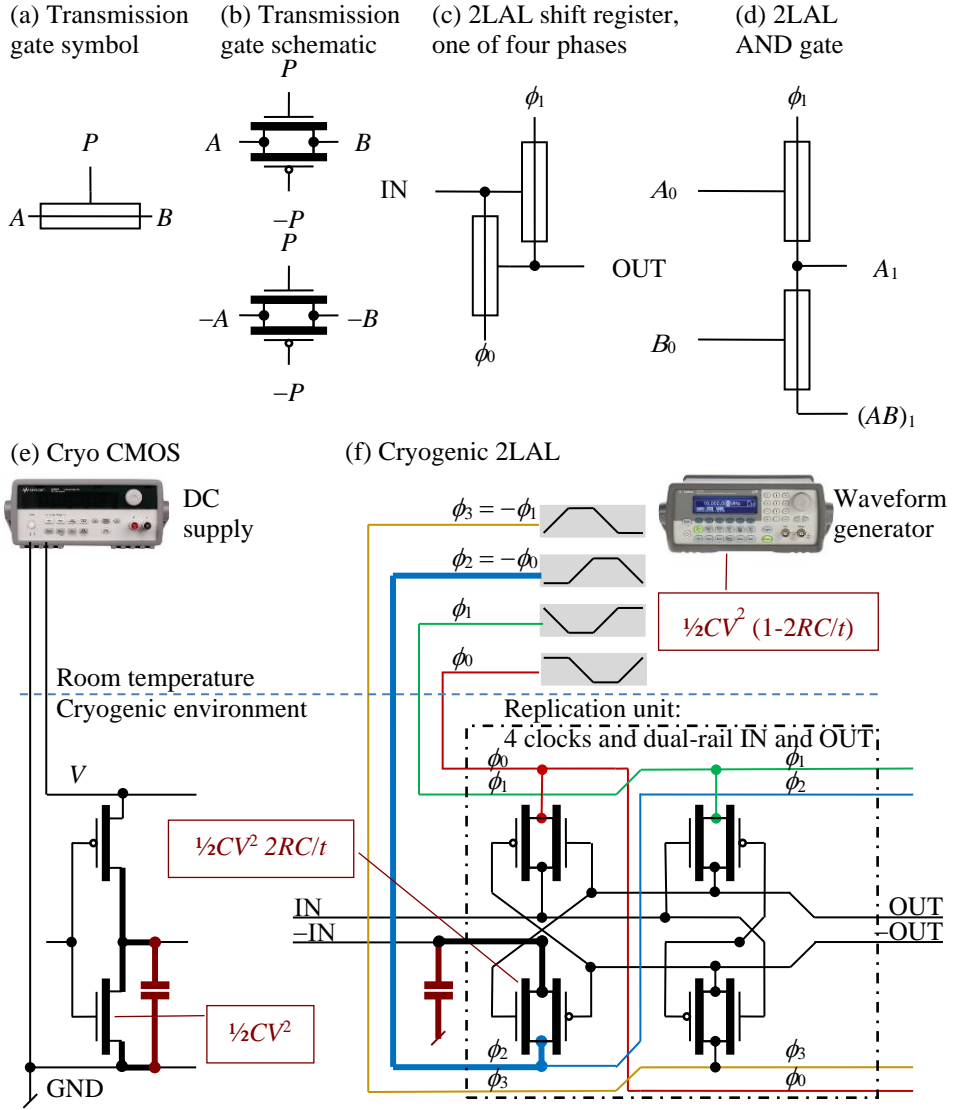

Fig. 3. (a-d) 2LAL circuits, see text. (e) Cryo CMOS setup; $\frac{1}{2}CV^2$ energy dissipated as heat in the transistor; incurring cooling overhead. (f) 2LAL setup; total of $\frac{1}{2}CV^2$ energy dissipated in the transistor and clock/power supply, of which only the transistor incurs a cooling overhead. Some figures inspired by Benjamin Gojman August 8, 2004, http://www.dna.caltech.edu/cbsss/finalreport/nanoscale_ind_gojman.pdf

### III. CRYOGENIC ADIABATIC TRANSISTOR CIRCUITS

#### A. Cryogenic operation makes adiabatic circuits practical—dynamic power

Both cryo CMOS and CATCs use voltage-based signaling on wires, which act as capacitors. The signal energy is $\frac{1}{2}CV^2$ in both cases, but the difference between the circuits is whether this energy is turned into heat in the cryogenic environment where it is subject to the refrigeration overhead or at room temperature where it is not.

Using the same data as Fig. 2b, Fig. 4[5] compares the power dissipated in TSMC 180 nm transistors when wired into CMOS and 2LAL[7] circuits, with the curves in each group representing different supply voltages, biasing strategies, and so forth. On a log-log scale, the curves show power dissipation of CMOS declines with slope −1 (linear) while 2LAL has slope −2 (quadratic).

As illustrated in Fig. 3e, a CMOS circuit charges and discharges wire capacitance from a DC power supply, dissipating ½$CV^2$ energy in the transistors' channels on each signal transition. Dissipation may occur as often as once per clock cycle, so the CMOS power curves in Fig. 4 are inversely proportional to the clock period, or slope −1. The red callout in Fig. 3e shows the location where the energy is transformed to heat. In a cryo CMOS system, this location is in the cold environment and therefore subject to refrigeration overhead.

An adiabatic circuit's combined clock and power supply waveforms have smooth ramps of d$V$/d$t \propto 1/t = 4f$, where $t$ is the length of the ramp and there are 4 ramps in a clock cycle. Given a low transistor source-drain drop, the charging current of the circuit's capacitance will be $I = C$ d$V$/d$t$ and the power dissipated in the transistor channel will be $I^2R$, where $R$ is the effective source-drain "on" resistance. As the clock period increases, $I$ falls linearly and $I^2R$ drops quadratically, this is why the 2LAL power curves in Fig. 4 decline with slope very close to −2.

In the limit of large $t$, the energy dissipated will be $C^2V^2R/t$. This expression can be algebraically rewritten as ½$CV^2$ $2RC/t$, which contains the recognizable energy term ½$CV^2$ and a multiplicative factor $2RC/t$. Energy is conserved, so the remaining ½$CV^2$ (1-2$RC/t$) on the capacitor has to go someplace. Per Fig. 3f, current from the red capacitor flows through a transistor, through the blue clock wire, and to the waveform generator. The red callouts in Fig. 3f show the two locations where energy is turned into heat, and the amount of heat created at each location.

In the original vision of adiabatic computing, an energy recycling power supply would collect the remaining ½$CV^2$ (1-2$RC/t$) energy from each clock cycle and use it for the next one, reducing wall plug power consumption. Unfortunately, energy recycling power supplies were never perfected and adiabatic systems built today would just move energy from the circuit to the power supply before turning it all into heat.

Even without an energy recycling power supply, the cryogenic implementation in Fig. 3f will bypass the refrigeration system for small values of 2$RC/t$ and save refrigeration energy. Examples later in this paper will use $t$ = 1,000 $RC$, leading to a 99.8% reduction of cooling overhead and wall plug power consumption.

Room temperature adiabatic transistor circuits have not been practical to date. Yet substituting cryogenic operation for the elusive energy recycling power supply yields a way to take engineering advantage of the unique characteristics of adiabatic transistor circuits.

## B. Extending the adiabatic speed range—static power

How can the range of the downward sloping region in Fig. 4 be extended? Let's first consider changes that do not involve a new fabrication process.

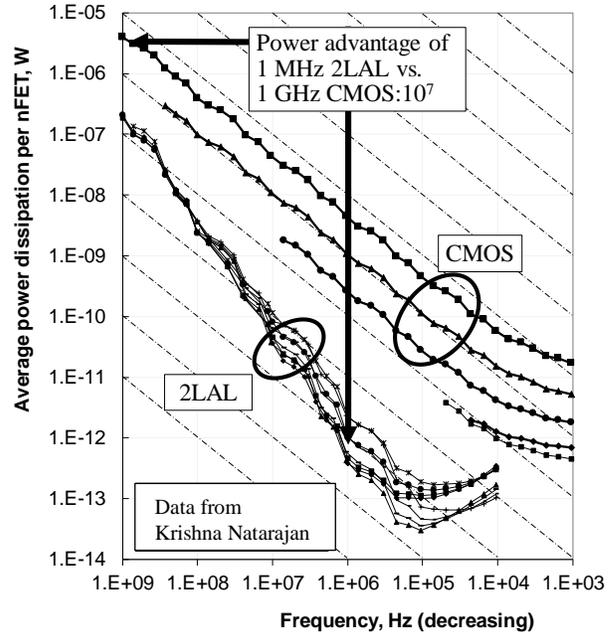

Fig. 4. Comparison of circuit efficiency for standard CMOS (top) and an adiabatic circuit 2LAL (bottom), showing a maximum advantage of 1,000× at 200 KHz. However, if 2LAL is operated at 4 K, down sloping curves should extend further, leading to a possible 100,000× energy efficiency improvement over room-temperature electronics. This may allow transistorized 2LAL to compete with JJs in applications where speed is not essential.

Most of the work in adiabatic computing has been for computing circuits, like microprocessors, where the designers accept the slow speeds in Fig. 4 but don't want their microprocessor to run any slower than necessary. However, this paper is about a hybrid technology where the designer can satisfy the need for speed with cryo CMOS or JJs. The purpose of the adiabatic circuits is to provide a solution for memory, state, or large amounts of logic where the value is in its complexity not its speed. These considerations give us a strategic reason to find the lowest possible energy for transistor circuits.

Generally speaking, the curves level off on the right of Fig. 4 at the point where transistors have the full power supply voltage across two terminals, causing either gate or source-drain leakage.

Transistors optimized for room temperature should benefit from cooling to some extent.[9] Total device leakage is the sum of temperature-independent gate leakage plus temperature-dependent source-drain leakage—where the two leakages can be traded off against each other by varying gate dielectric thickness. Assuming a fixed operating voltage, a wise process engineer should pick a gate dielectric thickness so that the gate and source-drain leakages are the same at room temperature, as shown on the left of Fig. 5.

Cryogenic operation causes a steepening of the subthreshold slope and a reduction in source-drain leakage,[2] as shown in center of Fig. 5, leaving gate leakage as the dominant factor in total leakage. If the leakages were balanced to begin with, the unchanging gate leakage should limit the adiabatic energy savings to about a factor of two, which is not enough to satisfy the needs of cold electronics.

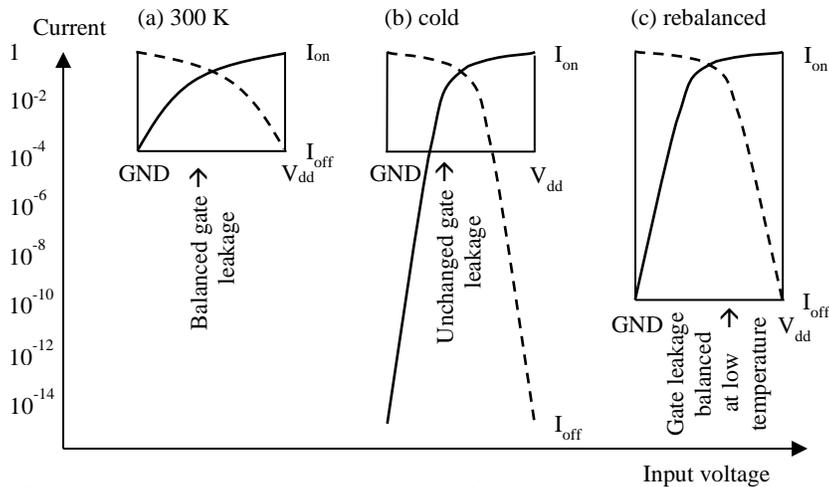

Fig. 5. Rebalancing transistors (a) Optimization for room temperature power dissipation suggests balancing source-drain and gate leakage. (b) Source-drain leakage drops significantly at cryogenic temperatures, but this will not make much of a difference. (c) However, the process can be optimized for cryogenic operation. Graphs show inverter current, nFET current is the solid curve and pFET current is the dashed curve.

It is easy to change the supply voltage, and a modest reduction will reduce gate leakage while increasing source-drain leakage, tending to bring the two into balance at a lower level of total leakage. While this is desirable, the amount of reduction in supply voltage is limited by the threshold voltage, which is not temperature dependent, so reducing the supply voltage should help but may not be the best solution.

The discussion above gives a new path to more energy efficient cryogenic memory-like circuits even without changing the physical structure of transistors. Both cryo CMOS and CATCs can be used for data storage, as illustrated by the shift register in Fig. 4. Transistors have the same leakage characteristics irrespective of whether they have been wired into a CMOS or a 2LAL circuit, so all the curves in Fig. 4 would have the same dissipation in the zero-frequency limit if they had been simulated with the same operating points.

The discussion above makes CATCs an effective memory option for some combinations of speed, power, and density. Logic is usually rated by speed and energy per operation, but memory can be useful holding data even if it does very few operations. Memory is also rated by density, so small devices are better.

CMOS SRAM will have somewhat fewer transistors than an adiabatic memory due to simpler circuits for address decoders and possibly cells. For the same transistors, CMOS should have an advantage at the lowest speed range. However, CATCs will have much lower energy per operation and will operate at lower power at even modest frequencies.

## C. Optimizing transistors

A process engineer can optimize transistors for a lower temperature by making the gate dielectric thicker than the room-temperature optimum until the gate and source-drain leakages are the same at the lower temperature.[9]

If total leakage could be reduced, the qualitative result should be extension of the region in Fig. 4 with downward slope −2 before $I_{off}$ and static leakage cause a leveling off. As of today, nobody knows how far leakage could be reduced if there were a deliberate effort to optimize transistors for that purpose.

## D. Adiabatic scaling of a CATC-RQL hybrid

The density and speed properties of CMOS and DRAM made them an especially effective hybrid for computer system design. Will CATCs and RQL form a hybrid that is similarly helpful?[10]

Adiabatic scaling refers to changing the clock frequency of an adiabatic circuit while simultaneously adjusting the number of gates on the chip so total chip power remains unchanged. For example, a chip with $g$ adiabatic gates operating at clock rate $c$ would have its clock rate lowered to $\alpha c$, for a scale factor $\alpha < 1$. This would cause each gate to dissipate $\alpha^2$ as much power, but in lieu of reducing chip power, the gate count would be increased to $g/\alpha^2$. As defined here, adiabatic scaling is similar to Moore's law. Moore's law is often defined by a factor $\alpha = 0.7$, which is the dimensional ratio between each process node and the previous one. This results in $1/\alpha^2 \approx 2\times$ as many transistors on a chip, with each chip expected to dissipate the same amount of power.

Adiabatic scaling is not "for free." Scaling causes the power supply leads in Fig. 3f to carry more current, requires lower and lower leakage transistors, and eventually fills up the chip so there is no space to hold more transistors.

Physical hybrids of JJs and cryogenic transistor circuits exist. JJ chips are fabricated in repurposed semiconductor fabs by evaporating a superconductor, such as Niobium, onto a blank silicon wafer. To manufacture a hybrid, the process just starts with a completed silicon wafer instead of a blank one.

Fig. 6 shows the hybrid's physical structure, with the beginning and ending points of adiabatic scaling based on the performance figures for a CMOS and RQL in Fig. 2a.

The projection of adiabatic scaling begins with the upper, superconductor layer filled with gates and the lower, semiconductor layer filled with enough cryo CMOS gates to make the dissipation of the two layers equal. Each subsequent scaling step has the same upper, superconductor layer but fills the lower, semiconductor layer with CATCs scaled adiabatically from the previous scaling step, which will leave the dissipation of the two layers equal.

TABLE I. TABLE 1: WORKSHEET ON ADIABATIC SCALING

| Baseline | | | |
|---|---|---|---|
| $N_{RQL}$ | $f_{RQL}$ | $P_{RQL}$ | $P_{Static}$ |
| 1 M | 1.6 GHz | 160 μW | n/a |
| $N_{CMOS}$ | $f_{CMOS}$ | $P_{CMOS}$ | $P_{Static}$ |
| 1 K | 4 GHz | 160 μW | n/a |
| A thousand extra gates, possibly useful for voltage-based signalling | | | |
| Scaling Step 1 | | | |
| $N_{RQL}$ | $f_{RQL}$ | $P_{RQL}$ | $P_{Static}$ |
| 1 M | 1.6 GHz | 160 μW | n/a |
| $N^{(1)}_{CATC}$ | $f^{(1)}_{CATC}$ | $P^{(1)}_{CATC}$ | $P^{(1)}_{Static}$ |
| 10 K | 400 MHz | 160 μW | 16.7 nW |
| Ten thousand slower gates, possibly useful for voltage-based signalling | | | |
| Scaling Step 2 | | | |
| $N_{RQL}$ | $f_{RQL}$ | $P_{RQL}$ | $P_{Static}$ |
| 1 M | 1.6 GHz | 160 μW | n/a |
| $N^{(2)}_{CATC}$ | $f^{(2)}_{CATC}$ | $P^{(2)}_{CATC}$ | $P^{(2)}_{Static}$ |
| 1 M | 40 MHz | 160 μW | 1.67 μW |
| Doubles gate count, but the new gates are slow | | | |
| Scaling Step 3 | | | |
| $N_{RQL}$ | $f_{RQL}$ | $P_{RQL}$ | $P_{Static}$ |
| 1 M | 1.6 GHz | 160 μW | n/a |
| $N^{(3)}_{CATC}$ | $f^{(3)}_{CATC}$ | $P^{(3)}_{CATC}$ | $P^{(3)}_{Static}$ |
| 100 M | 4 MHz | 160 μW | 167 μW |
| Similar resource mix to logic and memory, but also computes | | | |

Table I[*] takes the system in Fig. 6 through three adiabatic scaling steps of 10× clock period and 100× gate count. However, the first step switches the circuit design from CMOS to CATCs, the latter assumed to be 10× more complex, so the first increase in gate count increase will be 10× instead of 100×.

Scaling step 3 is of particular interest because it makes CATC a memory option. Scaling step 3 yields $N_{RQL}$ = 1 M fast RQL gates and $N^{(3)}_{CATC}$ = 100 M CATC shift register stages running at 4 MHz, which a designer will recognize as a resource mix similar to logic and memory.

However, CATC is actually a logic family, and this paper will describe ways of using CATC gates at scaling step 3 for important functions applicable to cold electronics applications.

Furthermore, the speed of gates in scaling steps 1 and 2 fit in between the speed of scaling step 3 gates and RQL, making them suitable for speed matching.

Let's pause to understand the historical context. Adiabatic circuits were studied in the 1990s,[8] but scaling step 3 would not have been reasonable at that time. At the micron linewidths of the day, chips could not hold 100 M gates and studies of

---

[*] The energy per operation $E$, propagation delay $t_{pd}$, and clock rate $f$ (assuming 500 gate delays per clock cycle) for RQL, CMOS, and CATCs from Fig. 2:

$E_{RQL}$ = 0.1 aJ, $t_{pd, RQL}$ = 1.25 ps, $f_{RQL}$ = 1.6 GHz

$E_{CMOS}$ = 40 aJ, $t_{pd, CMOS}$ = 0.5 ps, and $f_{CMOS}$ = 4 GHz

The IARPA Cryogenic Computational Complexity (C3) program created a million-gate RQL chip, so let's use $N_{RQL}$ = 1 M gates, so the superconductor layer will dissipate $P_{RQL} = N_{RQL} \times f_{RQL} \times E_{RQL}$ = 160 μW at 4 K, which corresponds to $N_{CMOS}$ = 1 K gates.

Superscripts [(1)] [(2)] and [(3)] indicate the scaling step.

The last column computes growing $P_{Static}$ power due to leakage. The leakage power calculation assumes a 1 V supply voltage, 3 KΩ on resistance, $I_{on}/I_{off} = 10^8$,[2] and a 50% duty cycle.

transistors at cryogenic temperatures revealed carrier freeze out and kinks in a key current curve. It is only due to recent interest in quantum computing that today's smaller transistors have been reassessed[2] and the disruptive effects no longer appear at smaller scale and different doping levels.

### E. Properties of a CATC-cryo CMOS hybrid

The properties of a CATC-cryo CMOS hybrid are also visible from Table I by ignoring the RQL layer on the blue background and assuming all the CATCs on the red background are intermixed on a chip.

Clearly, CATC's advantage in energy efficiency is so large that it should be used wherever possible. This would limit cryo CMOS to functions that do not have a parallel implementation or activities that require generation of high-speed signals.

Control systems for spin qubits would be one current application for a CATC-cryo CMOS hybrid. The hybrid may not support as many as $10^8$ qubits, but it could include interface electronics to a qubit-containing payload and some control electronics.

### F. Suitable memory-like structures

Today's commercial memories almost always allow random access, but the combination of fast random access and high density does not seem feasible at cryogenic temperatures. However, Fig. 7 shows a high capacity, high bandwidth memory-like structure that could be used for sequential storage.

The structure in Fig. 7 stores data in a serial shift register with 2,000-bit words built with gates from scaling step 3. This

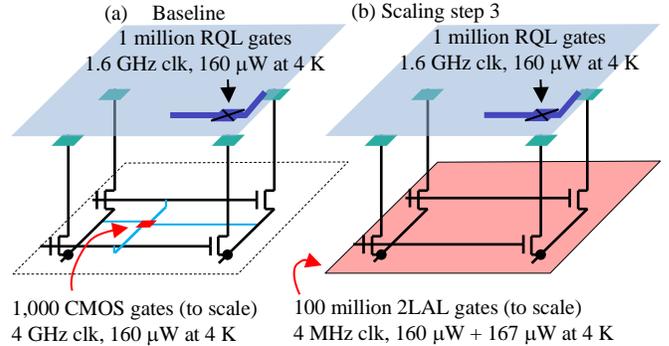

(a) Baseline
1 million RQL gates
1.6 GHz clk, 160 μW at 4 K

(b) Scaling step 3
1 million RQL gates
1.6 GHz clk, 160 μW at 4 K

1,000 CMOS gates (to scale)
4 GHz clk, 160 μW at 4 K

100 million 2LAL gates (to scale)
4 MHz clk, 160 μW + 167 μW at 4 K

Fig. 6. (a) With little or no adiabatic scaling, the semiconductor CMOS gates in red add process complexity but the number of gates isn't enough to make a difference to the design. (b) With the adiabatic scaling available at cryogenic temperatures, 100 million new semiconductor 2LAL gates dwarfs the original 1 million RQL gates and allows the module to address more complex problems.

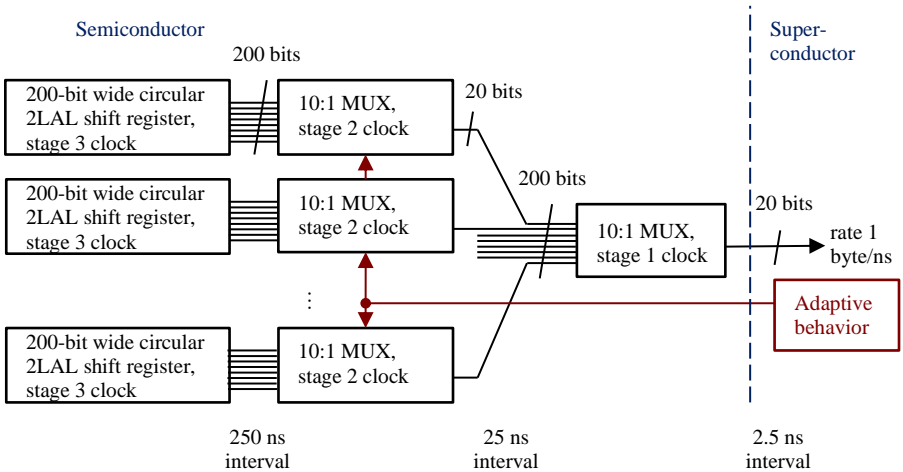

Fig. 7. CATC subsystem for sequential storage. To meet power requirements, the semiconductors must be slowed down to meet control signal requirements. However, the multiplexing scheme involving both the semiconductors and much faster superconductors will do the trick. The latency path in red is expected to be fast enough.

will allow sequential access at $f^{(3)}_{CATC}$ = 4 MHz, or 1 GB/sec. If the register loops back on itself, it can be loaded during the system boot process and the contents used many times. The shift registers may be as long as necessary to meet application storage needs, subject to chip size limitations.

However, transferring this data directly to RQL would require 2,000 receivers running at 1/1,250 of their maximum speed. To make more efficient use of resources, Fig. 7 shows 10:1 multiplexers using gates from scaling step 2 to create a 200-bit wide stream clocked at 40 MHz. A second level of multiplexers creates a 20-bit wide at 400 MHz.

Thus, the circuit in Fig. 7 has data density similar to memory but uses CATC's variable speed logic to process the data into a stream suitable for the much faster RQL logic.

### G. Energy efficient digital control signals

While others have proposed a circuit derived from DRAM for generating control signals in a cryogenic environment,[11, 12] this paper describes a more energy efficient approach using CATC addressing logic.

The lower layer in Fig. 8 shows DRAM memory cells that not only hold data for access from an external processor, but also "tap" each cell with a wire. The wire runs to another portion of the system carrying the state of the cell as a digital control signal.

Going beyond the current state of the art,[11, 12] the DRAM-derived circuit in Fig. 8 stores data on the capacitor plate at the interface between layers. The capacitor plate is actually the gate of a superconducting FET and the effect of the control signal is to control the critical current of a JJ.

By using CATC address decoders, the process for updating the control signals can be fully adiabatic, meaning the energy for an update could vary with speed according to the quadratic curve in Fig. 4, including the energy to charge the capacitive loads of the DRAM cells and the control signal.

The update process begins and ends in a reference state where all access transistors are in the off, or nonconducting, state and a copy of all the control signals are in the memory of an external processor.

Step 1. To update the programmable voltages on a row, the external processor transmits its copy of all the programmable voltages to the column data logic, which drives the control values to the source terminal of all the access transistors. The access transistors block further current flow because they are all turned off.

Step 2. The adiabatic row decoder translates the binary address from the external processor to a 1-of-$N$ signal that identifies the row, driving the signal to the gates of all the access transistors on the selected row. The natural operation of the adiabatic logic charges the transistor gates with very low dissipation, again following the quadratic curve in Fig. 4. There will be very little initial current flow through the transistors that turn on because the external processor used it's copy of the control signal data to drive each column with the same voltage as the control signal at the row-column intersection—i. e. each transistor's source and drain will be at the same voltage when the transistors turns on.

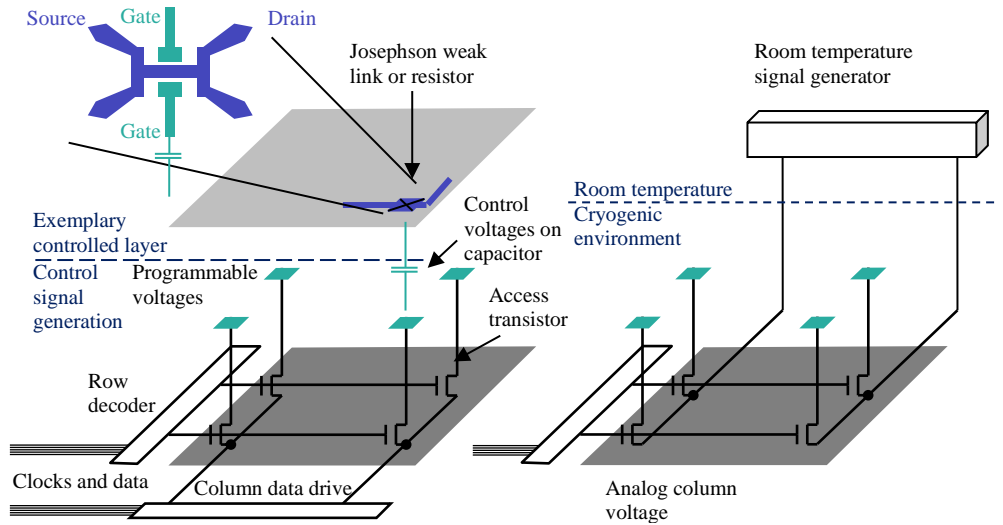

Fig. 8. A possible semiconductor-superconductor hybrid for control signal generation. The semiconductor layer applies voltage-based signals to the gates of superconducting FETs, which translate the signals into a form readily used by superconducting circuits. The hybrid would be fabricated by using a CMOS wafer as a base for depositing superconductor circuits—in lieu of today's method of using a blank Silicon wafer as a base.

Step 3. The external processor then transmits new data to the column data logic block. The natural operation of the adiabatic logic will charge or discharge the programmable voltages through the access transistors with very low power dissipation.

Step 4. The external processor then instructs the row decoder to turn off all access transistors. If the external processor retains the new signal values in its memory, the system will have been restored to the expected state between uses of this process.

The high energy efficiency of adiabatic circuitry comes with some unusual properties that must be considered but are not new.

A four-phase clocked logic family has been developed around 2LAL, which includes a signaling specification that requires each data signal to be valid during one of the clock phases. While a string of 0s in 2LAL produces a DC value at the clock's low voltage $V_L$, a string of 1s produces an AC signal that transitions between $V_L$ and the clock's high voltage $V_H$. The latter signal meets the signaling specification, but is in different states at other times. This behavior is transparent when connecting 2LAL gates to each other, but the DRAM access transistors are not 2LAL gates so the complete signaling behavior must be considered.

The access transistors in Fig. 8 require certain voltages to function properly, such as source-gate voltages that reliably turn the transistor on or off. The row decoder and column drive circuits will be driven by two separate sets of 2LAL clocks of the general form shown in Fig. 3. However, the external clock generator determines $V_L$ to $V_H$, which can be different for the two functions. For Fig. 8, the row decoder's waveforms would swing from $V_{L,row}$ to $V_{H,row}$ and the column data drive waveforms would swing from $V_{L,data}$ to $V_{H,data}$. A transistor in the on state would see a source-gate voltage of $V_{H,row} - V_{L,data}$, and similarly for the off state. Proper engineering of these voltages would probably lead to two sets of four combined clock and power supply signals.

This paper uses 2LAL as an example, but the ideas apply to other logic families, such as SCRL.[8] Instead of following the 2LAL convention of 0s being a DC level and 1s being a signal between $V_L$ and $V_H$, SCRL signals return to an intermediate value $(V_L + V_H)/2$ during certain phases of the clock. Other adiabatic logic families may have their own requirements.

### H. Analog control signals, DC or AC

This paper describes how to create analog control signals, as illustrated on the right side of Fig. 8. While adiabatic principles apply to analog signals, there isn't a general way of extending an adiabatic logic family like 2LAL to handle analog signals. However, the analog signals only appear on the columns, so only the column data drive portion of Fig. 8 needs attention.

In lieu of an adiabatic digital column driver, the approach is to run a wire for each column to an analog voltage generator at room temperature. If the voltage generator follows the protocol for column data driver, the energy efficiency would follow the quadratic curve in Fig. 4.

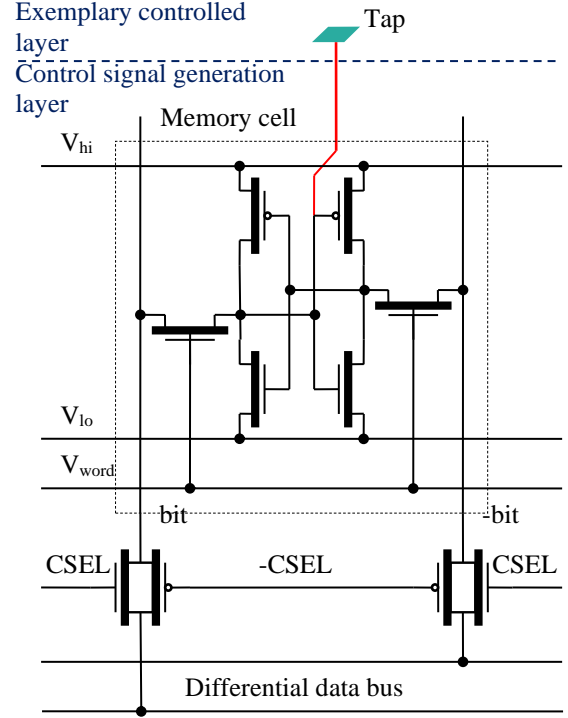

Fig. 9. A purpose-built adiabatic memory similar to conventional SRAM, with taps. The main cell comprises 4 transistor in a cross-coupled inverter configuration and two access transistors. Unlike conventional SRAM, the cell's power comes from row and column wires. Each cell is tapped.

However, other applications may require high speed analog signals, such as the high-speed pulses in Ref. 12. The high-speed pulses in this situation will pass through resistive transistor channels and dissipate power, but the approach in this paper nonetheless improves the energy efficiency of the row decoding.

### I. Adiabatic SRAM alternative

An alternative approach is to use the adiabatic equivalent of an SRAM[13] in lieu of the DRAM just discussed. The adiabatic SRAM circuit replaces a single access transistor per bit, as shown in Fig. 9, with four transistors in a cross-coupled inverter configuration, connected between two floating power supplies. The bit cell includes two additional access transistors.

This paper will just summarize the baseline operation of an adiabatic SRAM. The CMOS addressing logic in a standard SRAM operates at the power levels of baseline level in Table I, which is strong enough to overpower the cross-coupled inverters in the storage cell. An adiabatic SRAM uses adiabatic logic for address decoding, which reduces power quite a bit. However, overpowering the single digital signal in the memory cell would create more heat than the rest of the memory, at least at the speed of scaling step 3 of Table I. Since the power in the adiabatic memory comes from the row and column drive, individual cells can be essentially powered down adiabatically, switched, and then powered up adiabatically in a new state. See Ref. 13 for additional detail.

In Fig. 9, each SRAM storage cell is tapped as indicated.

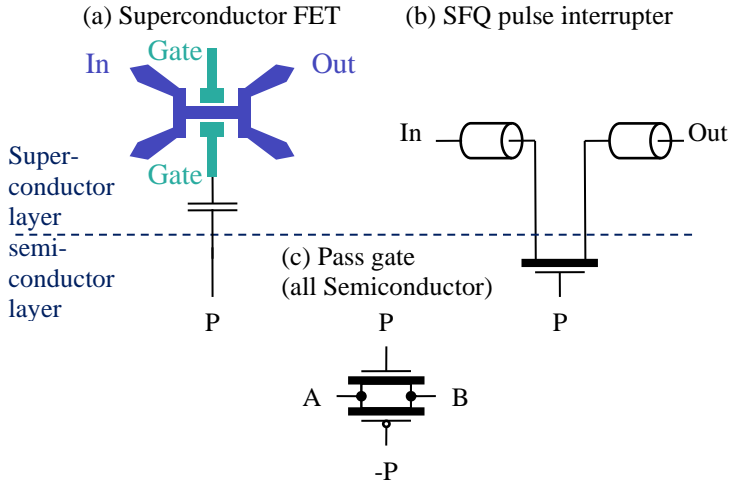

Fig. 10. Options for transferring control signals between signal forms of the two technologies. (a) Control voltages pass between layers and then influence a (currently experimental) superconductor FET, (b) an SFQ pulse generated in the superconductor layer passes to the semiconductor layer via an ohmic (non superconductor) wire, passes through a large, but otherwise ordinary, transistor, and back to the superconductor layer. If the transistor is on, the SFQ pulse becomes somewhat attenuated, if off, the SFQ pulse is almost entirely blocked (c) pass gate for controlling other parts of the semiconductor layer.

## J. Use of control signals

Fig. 8 and Fig. 9 illustrate how either a DRAM-type control signal can influence a JJ circuit on another layer, with Fig. 10 providing more detail and two other options. Both DRAM- and SRAM-generated signals provide a voltage that can influence other parts of the system through an electric field.

Fig. 10a illustrates a superconducting field-effect transistor (FET).[5, 14, 15] Ignoring the green structures for the moment, the blue structure is a superconducting wire that will conduct current horizontally with zero resistance. However, a narrow superconducting wire only conducts with zero resistance up to a maximum current, called the critical current, above which the device becomes a resistor.

Superconductivity can be disrupted by an electric field, such as the field due to the programmable voltage across the green capacitor in Fig. 8 or Fig. 9. Theory and experiment for the superconducting FET show the weak link's critical current changes when the green structure applies a few volts or more, positive or negative. While current demonstrations[14] required a higher voltage, there is an expectation that the drive voltage could be as low as 2.5 V, a reasonable voltage swing for transistorized circuits. This could lead to a structure like shown in Fig. 8, where CMOS voltage-based signals are converted to the single-flux quantum (SFQ) signals typical in JJ circuits.

Fig. 10b illustrates a large but otherwise standard semiconductor FET in a role where it can interrupt a superconducting SFQ pulse. SFQ pulses propagate efficiently along transmission lines, which have a characteristic impedance of around 15 Ω for Niobium superconductor chips at 4 K, leading to SFQ pulse dimensions of about 1 mV × 2 ps. If such an SFQ signal is routed through a large semiconductor FET, with an on resistance of around 15 Ω, the pulse will pass with some attenuation. If the transistor is off, it will not pass at all.

Thus, a control signal can influence the JJ circuit by blocking or passing an SFQ pulse. The energy consumed is just the energy in the pulse, if the pulse is destroyed or attenuated.

Fig. 10c illustrates a control voltage influencing more semiconductor circuitry. If the control voltage is applied to the gate of a semiconductor FET, the FET can act as an SPST switch, with the control voltage shorting the source and drain or leaving an open circuit. The CMOS transmission gate illustrated has better properties, but requires two transistors and complementary control signals.

## IV. DYNAMICALLY RECONFIGURABLE CRYOGENIC FPGA

This paper describes two variants of a cryogenic FPGA, both using CATC control signals to configure the FPGA's programmable logic.

In the first variant, the programmable logic is RQL on a separate layer with the principle advantage that valuable space on the superconducting layer is not taken up with configuration logic unnecessarily.

In the second variant, the programmable logic is cryo CMOS on the same layer, with the advantage of higher energy efficiency during reconfiguration.

An FPGA generally comprises an array of configurable logic blocks (CLBs) connected by a programmable routing network, as shown in Fig. 11. The FPGA simulates an integrated circuit by configuring each CLB to be the equivalent of a few gates. The routing network is configured to replicate an integrated circuit's netlist. Just as memories are manufactured without any data, FPGAs are manufactured without a specific function. A configuration string sets the FPGA's function during the boot process.

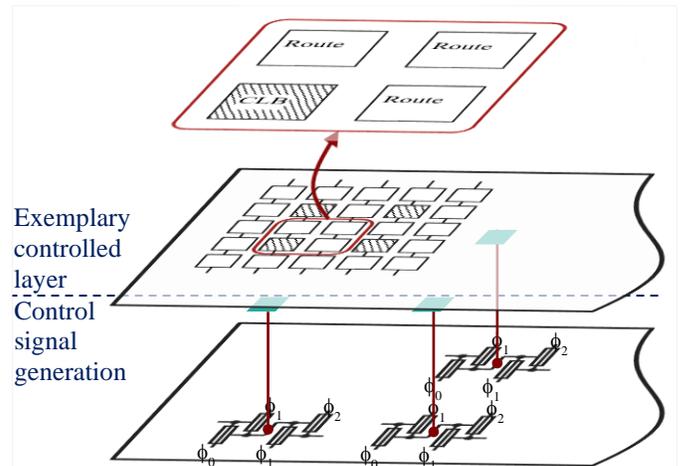

Fig. 11. Basic structure of a hybrid FPGA comprising configurable logic blocks (CLBs) connected by a network of routers, whose overall routing pattern is controlled by setting each open circle to route data between its inputs and outputs in a specific pattern. Control signals give each CLB a specific identify and set the routing pattern to duplicate a logic deisgn. This structure has two layers, one for the configuration logic and another for the configured logic, thus benefitting from the density (complexity) of the semiconductor configured logic and the high-speed/low-power characteristics of the JJ-based configured logic.

For example, a CLB could support Boolean AND, OR, NOT, and a half adder, with two control signals selecting one of the four functions. Likewise, control signals for routers could specify whether data continues in the same direction, turns left, turns right, or connects to the nearest gate. In both cases, the control signals would be generated by adiabatic transistor logic. If the programmable logic is RQL, the voltage-based signals would be transformed to SFQ via the structures in Fig. 10.

CMOS FPGAs have bidirectional pass gates, yet JJs are not easily configured to pass signals in both directions. As a consequence, superconductor FPGAs[16] use only unidirectional connections, but require more of them, resulting in higher overhead than equivalent CMOS FPGAs.

A more recent article[17] proposes creating superconductor FPGAs using a new magnetic JJ (MJJ) as the underlying programmable device. An MJJ has an internal magnet whose field can point in one of two directions. The MJJ's internal state causes its critical current to change somewhat, effectively disabling circuits that depend on a specific critical current. Selective disablement is the method influencing the configurable logic to create the desired function.

The first variant of the cryogenic FPGA combines the control signal circuit in Fig. 8 or Fig. 9 with the superconductor FPGA,[17] replacing the MJJs with the superconductor FETs. The superconductor FETs are not essential and the transistorized SFQ interrupter in Fig. 10b could be used instead.

## V.  QUANTUM COMPUTER CONTROL ELECTRONICS

Fig. 1 illustrates the physical form of today's superconducting quantum computers, except that control signals are routed between qubits and room temperature electronics with only passive processing along the way. Signal generation, analysis, and decisions are made by room temperature electronics with intermediate temperature stages being used for thermal sinks and noise attenuation.

The amount of wire carrying these signals across the temperature gradient grew as quantum computers scaled up even to today's modest levels, leading to space congestion in the cold environment and heat and noise flow through the wire from room temperature to the cold environment. Heat removal eventually overloads the cryogenic refrigerator and blocks further scaling. There seems to be a consensus based on Rent's rule[1] and other principled arguments that continued scaling will require control electronics in the cold environment, using some variant of the structure in Fig. 1.

Distributing the control function across multiple temperatures is a current research issue, one that CATCs can address. For example, only passive analog devices and digital multiplexers can be placed at the coldest temperature stages due to fundamental technology limitations. Digital controllers are necessary, but they are only viable at 4 K or higher. This leads to innovative designs where the controller is partitioned both functionally and across different temperatures[18, 19] to meet limitations on devices, materials, and architectures.

### A. *Control of spin qubits*

Quantum computers based on spin qubits also use the structure in Fig. 1, but scale up plans suggest something more elaborate. Spin qubits are electrons loosely bound to a location in a material, such as through a donor atom or a quantum dot. While experimental demonstrations have not gone beyond two qubits, there are published papers proposing designs up to $10^8$ qubits.

The designs for scaled up spin qubit systems have a qubit layer and a cryo CMOS layer. Spin qubits need low temperatures and cryo CMOS dissipates a lot of heat, so there is a potential heat problem. Published papers show that heat removal is possible, but do not include quantitative estimates. One architectural approach has a handful of wires between each qubit and the control electronics,[11, 12] while another approach comprises a 2D array of qubits with modest number of wires per row, column, or diagonal.[20] These approaches could be impractical at the scale of $10^8$ qubits, opening the possibility that the energy efficiency benefits of the hybrid could help.

Each qubit type requires control signals with certain properties, such as DC, AC, microwave, various noise levels, and so forth. For example, a recent survey[11] identified physical I/O requirements for quantum dot qubits as:

1. an independent DC voltage on every qubit (site) up to ±1 V

2. an independent voltage pulse with sub-ns rise times on every qubit up to tens of mV

3. an independent microwave magnetic or electric field at every site, typically −40 to −20 dBm, 1–50 GHz bursts of 10 ns to 1 μs duration

The signal types above have been addressed previously.

### B. *Control of superconducting qubits*

Superconducting qubits can be controlled with SFQ pulses directly.[19] This approach matches the capabilities of the hybrid very elegantly, allowing reconfigurable FPGA logic to create SFQ pulses that interact with qubits directly and with no per-qubit wiring to room temperature.

Traditionally, superconducting qubits have been controlled by shaped microwave pulses, or a microwave sine wave within a lower-frequency envelope waveform. Fig. 12 shows how to generate these pulses using the hybrid, where slower but more complex transistor circuits control a smaller number of high-speed analog microwave components built from JJs.

The first step is to store digitally encoded waveforms in the memory-like structure illustrated in Fig. 7 and transfer it to the RQL layer.

The next step is to convert the digitally encoded waveform into an analog signal amenable to these microwave components, which is a typically a current. Current sources controlled by SFQ pulses are available.[21]

The final step is to use one of the microwave switches, modulators, or other components developed by the quantum computer community for controlling microwave signals with currents,[22] which are typically generated by current sources at room temperature and transported through the temperature gradient on a microwave transmission line.

Fig. 7 also shows a feedback path from the high-speed electronics, allowing behavior in the payload to influence waveforms.

*C. Architecture of a cold, scalable controller*

This paper has now discussed all the components needed to create a complete system that addresses the issues described in Fig. 1. The controller will be described as a CATC-RQL hybrid, yet the ideas apply to a CATC-cryo CMOS hybrid as well.

The controller should be capable of generating complex control sequences at high speed and with low power. This paper uses a transmon quantum computer controller as an example, where the controller needs to produce control sequences for calibration, qubit initialization, quantum computer arithmetic, and qubit readout.

While Fig. 2 shows that RQL meets the speed and power requirements and Fig. 10 shows how to generate control signals, RQL gates are about 10,000× the size of CMOS gates, thus limiting the controller's scalability.

To get around the scalability limit, the CATC-RQL hybrid organizes the RQL logic into an FPGA as illustrated in Fig. 11, or another reconfigurable structure. The RQL logic is then reconfigured on the fly to produce different behaviors sequentially, thus increasing the apparent number of RQL gates by a principle similar to timesharing.

However, the FPGA behaviors should switch without the qubit control signals stopping during reconfiguration and stalling the overall system, which would not only waste time but may allow the system's state to deteriorate, such as qubits decohering.

A configuration buffer reduces the possibility of stalls, illustrated in Fig. 13a as a $k$-bit wide by 4-stage cyclic 2LAL shift register. The number $k$ corresponds to the number of FPGA configuration bits and the buffer's $k$-bit output is used as a form of tapped memory to configure the FPGA.

Let us go through a sample operating sequence, putting in rough timing numbers:

1. After power-on, the external processor loads the 4 $k$-bit configuration sequences into the 4×$k$-bit configuration buffer shown in Fig. 13a. This could be done serially and should take less than a few seconds. The configuration buffer is shifted so the FPGA configuration for calibration appears on the $k$ outputs, leaving the RQL FPGA ready to calibrate the transmons, now shown in Fig. 13b.

2. The RQL clock is turned on, at perhaps 5 GHz, and generates the calibration sequence until the external processor decides to turn off the RQL clock.

3. The external processor commands the clock generator in Fig. 3e to create four phases of the 4 MHz combined clock and power supply, rotating the contents of the configuration buffer in Fig. 13a and b in 250 ns so the configuration for qubit initialization is transmitted to the superconductor layer.

4. The RQL clock is turned on and performs qubit initialization, which takes perhaps 5 μs, after which the external processor turns off the RQL clock.

5. The external processor commands the clock generator to shift the configuration buffer again, loading the quantum computer arithmetic configuration in 250 ns.

6. The RQL clock is turned on for perhaps 100 μs, or however long the qubits can operate without undue risk of decohering.

7. The external processor shifts to the readout configuration, performs readout, and the process completes.

This controller is viable because the timings are in the right proportions to each other and the qubit properties. While the CATCs must run slowly due to the slow speed of the gates in scaling step 3 of Table I, the architecture proposed can carry out an FPGA reconfiguration in 250 ns. While 250 ns is a long time compared to the 200 ps clock period of RQL, it is much shorter than the decoherence time of current transmons. Thus, the approach in this paper should be viable today, but will work better in the future as transmon coherence times improve.

However, Fig. 13a also includes a feedback path (labeled "alternative branch") for adaptive control. Each configuration of the controller allocates a small amount of RQL logic to detect conditions that require a complex response. For example, a quantum error detection and correction circuit will usually conclude that there has been no error. In the improbable but important case that an error is detected, the error correction process may be complex enough to require reconfiguration of the FPGA to generate a completely different control sequence. So, the RQL alternative branch signal can force a change in the shift register without direct involvement of the external processor. A more sophisticated implementation might allow a jump to a configuration outside the normal rotation pattern,

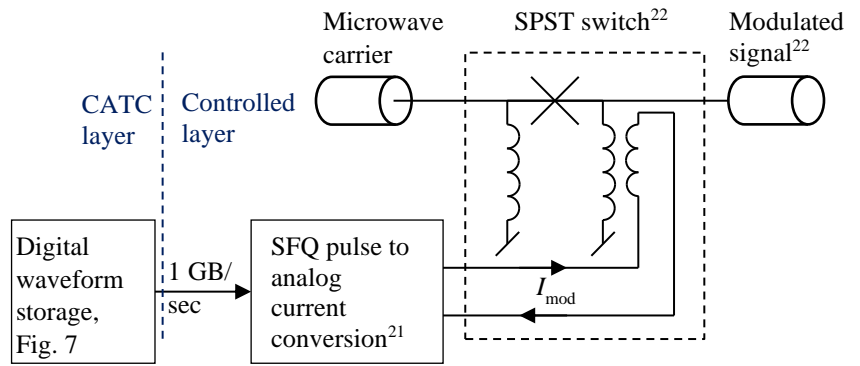

Fig. 12. Controlled microwave modulation. Pulse envelopes stored in the memory-like shift register in Fig. 7 are speed matched to RQL's faster clock. The gigabyte/second stream provides the control signal to a superconductor D-A converter that generates analog currents,[21] $I_{mod}$. The digitally controlled current becomes the flux control for an SPST microwave switch.[22] The superconducting transmon research community has mixers, phase shifters, and other circuits.

giving the FPGA reconfiguration some of the capability of a standard computer's branch instruction.

Due to the slow speed of CATC, the configuration buffer must be able to reconfigure the FPGA in just one or a few clock cycles. It could be a shift register of different dimensions, a structure with an access pattern different from a cycle, or multiple configuration buffers running independently.

## VI. VARIATIONS AND GENERALIZATION

In addition to the four exemplary quantum computer FPGA configurations, configurations could be created for different quantum error correction codes, such as 5-bit, 7-bit, or surface codes. This would allow a quantum computer to support any quantum error correction code without changing the hardware.

However, the same controller architecture could apply to a cryogenic sensor array that identifies extrasolar planets via a control sequence in the FPGA. The FPGA configuration could change as more is known about a potential planet, or as improved algorithms are developed.

The paragraph above also applies to subroutines in either classical or quantum algorithms. One algorithm might use 8-bit integer data types whereas another might use 150-bit integers. In fact, a single algorithm might use integers of several word sizes. A control sequence could be developed for each different integer size and loaded into the FPGA as needed.

This paper used CMOS (CMOS HP) and RQL as technology examples for the hybrid because their parameters were readily available in Fig. 2. However, the unlabeled dots represent other Beyond CMOS devices that could be looked up in Ref. 4 and considered for cryogenic operation. There are also other circuit families that could have the same qualitative behavior as CMOS and 2LAL such as Split-Level Charge Recovery Logic (SCRL), Efficient Charge Recovery Logic (2N2P or ECRL), 2N2N2P, Positive Feedback Adiabatic Logic (PFAL), Differential Cascode Pre-resolve Adiabatic Logic (DCPAL), and others. There are other purpose-built adiabatic memories as well.[13] JJs are a device that is the principal component in cryogenic logic that signals with SFQ pulses, of which Rapid Single Flux Quantum (RSFQ) is the historical example, but Energy-efficient RSFQ (ERSFQ), eSFQ, and Reciprocal Quantum Logic (RQL) are variants. There may be others.

This paper used three stages operating at 300 K, 4 K, and 0.015 K as an example, but the ideas apply to two or more stages at temperatures down to a ratio of about 10:1 between the warmest and coldest. The ideas should still work even at lower temperature ratios (e.g. 2:1), although the benefits decrease as the temperature range approaches a single temperature. The term "room temperature" is used to represent the approximate temperature of earth's environment, which is the heat bath for terrestrial systems. However, the ideas will apply even if the heat bath is at a lower temperature, such as space, or higher temperature, such as under the Earth's surface.

(a) Load mode

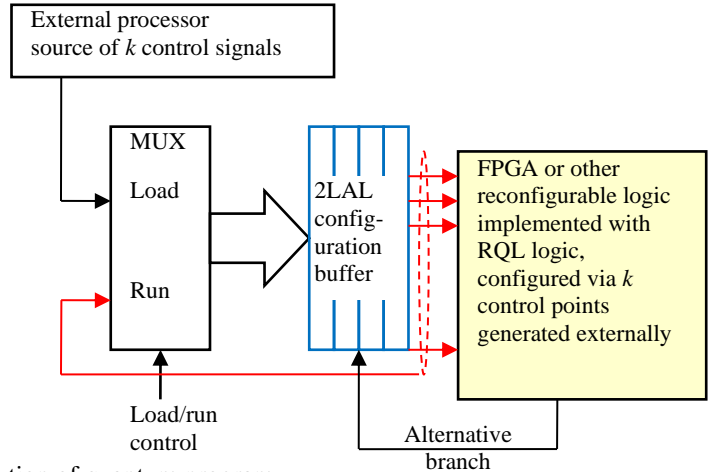

(b) Execution of quantum program

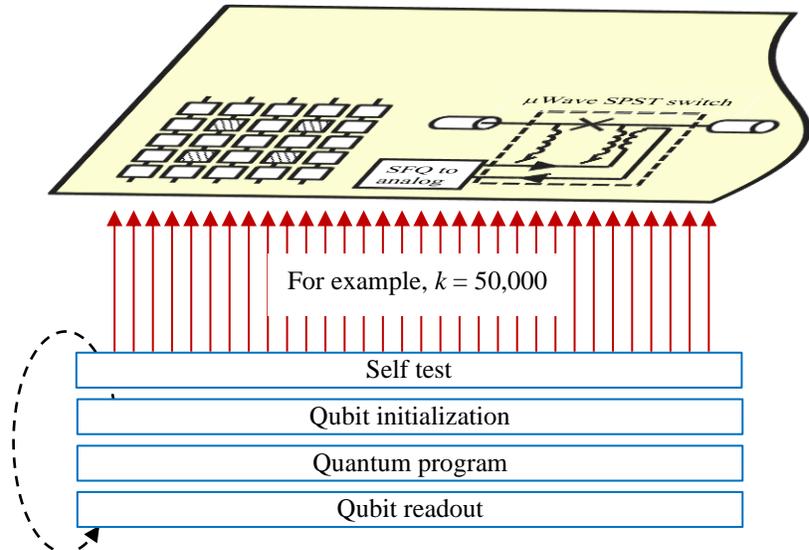

Fig. 13. Rapid reconfiguration example. A system has four modes of operation, each of which is specified by an FPGA with $k$ configuration bits or control points. An external controller directs the source of control signals to create four sets of $k$ control signals sequentially. With the multiplexer set to the select the load input, the four sets are loaded into the $k$-bit wide, four-stage shift register. Switching the multiplxer to the run position, each clock of the shift register exposes a the FPGA reconfigurable logic to the next mode, in rotating sequence. While the 2LAL shift register must be clocked at 4 MHz for energy efficiency, this is fast enough to completely reconfigure the FPGA within the decoherenance time of a qubit. As an option, the path labeled potential branch could convey information from the controlled payload that forces a reconfiguration, including to a configuration out of the normal rotation pattern, thus giving reconfiguration some of the capability of a standard computer's branch instruction.

VII. CONCLUSIONS

I've suggested a step forward based on the confluence of several trends. Cryo CMOS was always known to have a steeper subthreshold slope, but carrier freeze out and kinks in a key device curve made it hard to exploit the steepness. However, freeze out and kinks have almost disappeared due to the natural advance of Moore's law and transistors are more suitable for cryogenic operation than they used to be.

Since the performance of quantum computers comes from the qubits, not the control electronics, the single-minded pursuit of high-throughput transistors is less important, creating an opportunity for slower devices.

These two natural changes addressed the reasons adiabatic circuits did not catch on years ago. Substituting the ability to bypass the cryogenic refrigerator and it's overhead for the elusive energy-recycling power supply makes adiabatic circuits practical, allowing them to become a type of mortar that can bind bricks like classical JJ logic and qubits together into a complete system.

Using the technology base above, I've outlined a new approach for quantum computer control electronics. Von Neumann processors may be essential for controlling a quantum computer, but that doesn't mean the von Neumann processors have to be in the cold environment. For extreme quantum computer scale up, the cold environment will have to be reserved for those functions that cannot be performed anywhere else, such as low-level signal control, such as microwave modulators, and control functions that Rent's rule forces to be close to the qubits to avoid interconnect congestion in connecting to higher levels of the system.

The paper illustrates the new design principles with a semiconductor-superconductor hybrid quantum computer control system, where a memory-like shift register is combined with fast FPGA-like logic in a way that meets the requirements of Rent's rule for scaling.[1]

The advantages of hybrid CATC-JJ quantum computer control electronics would be in scalability. In lieu of simply proposing an architecture to be built out of a single well-known gate-level technology, such as cryo CMOS or classical JJs (an SFQ logic like RQL), I devised a hybrid of the two well-known technologies, except that the transistor technology uses a different logic circuit that has even lower power. The resulting classical electronics has a good balance for the quasi signal processing needs of quantum computer control electronics, and should scale further that previous solutions due to lower power.

As a next step for physical realization, growth of the Internet of Things led key semiconductor manufacturers to design processes with extremely low leakage, specifically Intel 22FFL, GF 22FDX, TSMC 22ULP, and ST 28 FDSOI. These would be good candidates for trial implementations of a CATC-JJ hybrid.

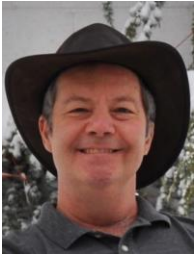

**ERIK P. DEBENEDICTIS** (S'18) received B. S. and Ph. D. degrees in Electrical Engineering and Computer Engineering from Caltech in 1978 and 1983, and an M. S. degree in Computer Engineering from Carnegie Mellon in 1979. He worked at Bell Labs, Ansoft (now ANSYS), nCUBE, NetAlive, Sandia National Labs, and is currently operating Zettaflops, LLC. His interests are in supercomputing, advanced computer technology, and quantum computing. He is the Editor in Chief of IEEE Transactions on Quantum Engineering and a co-lead of the IEEE Quantum Initiative.